\documentclass[conference]{IEEEtran}
\IEEEoverridecommandlockouts
\usepackage{cite}
\usepackage{amsmath,amssymb,amsfonts}
\usepackage{algorithmic}
\usepackage{graphicx}
\usepackage{textcomp}
\usepackage{xcolor}
\def\BibTeX{{\rm B\kern-.05em{\sc i\kern-.025em b}\kern-.08em
    T\kern-.1667em\lower.7ex\hbox{E}\kern-.125emX}}
\usepackage{amsmath, amssymb, bm, cite, epsfig, psfrag}
\usepackage{graphicx}
\usepackage{soul} 
\usepackage[margin=0.7in]{geometry}
\usepackage{dblfloatfix}
\usepackage{array}
\usepackage{lipsum}
\newcolumntype{P}[1]{>{\centering\hspace{0pt}}p{#1}}
\newcolumntype{M}[1]{>{\centering\hspace{0pt}}m{#1}}
\newcolumntype{L}{>{\centering\arraybackslash}m{3cm}}
\usepackage[font=small]{caption}
\usepackage{epstopdf}	
\usepackage{longtable}
\usepackage{supertabular,booktabs}
\usepackage{bbm}
\usepackage{multirow}

\usepackage{subfigure}
\usepackage{etoolbox}
\usepackage{pbox}
\usepackage{fixltx2e}
\usepackage{tabu}	
\usepackage{filecontents}
\usepackage{enumerate}
\usepackage{textcomp}

\usepackage{colortbl}
\usepackage{fancyhdr}

\usepackage{bm}
\newtoggle{conference}
\togglefalse{conference} 
\fancypagestyle{firststyle}{
	\fancyhf{}
	\fancyhead[L]{S. Ju, and T. S. Rappaport, ``Millimeter-wave Extended NYUSIM Channel Model for Spatial Consistency,''  2018 \textit{IEEE Global Communications Conference (GLOBECOM)}, Abu Dhabi, UAE, Dec. 2018, pp. 1-6.}     

}

\begin{document}

\title{Millimeter-wave Extended NYUSIM Channel Model for Spatial Consistency}

\author{\IEEEauthorblockN{Shihao Ju and Theodore S. Rappaport}
\IEEEauthorblockA{\textit{NYU WIRELESS}\\
\textit{NYU Tandon School of Engineering} \\
Brooklyn, NY, 11201 \\
\{shao, tsr\}@nyu.edu}
}

\maketitle
\thispagestyle{firststyle}
\begin{abstract}
Commonly used drop-based channel models cannot satisfy the requirements of spatial consistency for millimeter-wave (mmWave) channel modeling where transient motion or closely-spaced users need to be considered. A channel model having \textit{spatial consistency} can capture the smooth variations of channels, when a user moves, or when multiple users are close to each other in a local area within, say, 10 m in an outdoor scenario. Spatial consistency is needed to support the testing of beamforming and beam tracking for massive multiple-input and multiple-output (MIMO) and multi-user MIMO in fifth-generation (5G) mmWave mobile networks. This paper presents a channel model extension and an associated implementation of spatial consistency in the NYUSIM channel simulation platform \cite{Sun17a,Ju18a}. Along with a mathematical model, we use measurements where the user moved along a street and turned at a corner over a path length of 75 m in order to derive realistic values of several key parameters such as correlation distance and the rate of cluster birth and death, that are shown to provide spatial consistency for NYUSIM in an urban microcell street canyon scenario. 
\end{abstract}
    

\section{Introduction}~\label{sec:intro}
Recent work reveals that global mobile data consumption will experience a vast increase over the next few years \cite{Hemadeh18a,Rap13a}. MmWave communication is regarded as a promising technique to support the unprecedented capacity demand because of the availability of ultra-wide bandwidths. Accurate channel modeling for mmWave frequencies has been an important area of study recently, since the mmWave channel, when combined with directional antennas, has vastly different characteristics from omnidirectional microwave channels\cite{Rap15a,Sun18a,Rap15b}. Many statistical and deterministic channel models such as METIS \cite{METIS15a}, NYUSIM \cite{Sun18a,Samimi16a}, MiWEBA \cite{MiWeba14a}, 3GPP\cite{3GPP.38.901,3GPP-TR.25.966}, 5GCM \cite{A5GCM15}, and mmMAGIC \cite{mmMAGIC17a}, have been proposed over the past few years. 

Most of the existing statistical channel models are drop-based, where all parameters used in one channel realization are generated and used for a single placement of a particular user. Then, a subsequent simulation run of the drop-based channel model results in an independent sample function for a different user, and at a completely different, arbitrary location, even if the same distance between the transmitter (TX) and receiver (RX) is considered \cite{Sun17a,Sun18a,Shafi18a}. Drop-based models are popular because of their simplicity in Monte Carlo simulations\cite{Tranter03a}. The NYUSIM channel model generates static channel impulse responses (CIRs) at a particular distance/location, or across the manifold of a 2-D antenna structure, but cannot generate dynamic CIRs with spatial or temporal correlation based on a user's motion within a local area\cite{Sun17a,Sun18a,Samimi15a}. In other words, CIRs of two closely spaced locations are generated independently, although one would expect the CIRs to be highly correlated if the users were truly close to one another \cite{Shafi18a}. It stands to reason, and is borne out by measurements, that two close users, or a user moving in a small area, should experience a somewhat consistent scattering environment\cite{Ju18a}. Thus, spatial consistency has become a critical modeling component in the 3GPP Release 14 \cite{3GPP.38.901}. Challenges exist for drop-based models to be spatially consistent, since nearly all temporal and spatial parameters would need to vary in a continuous and realistic manner as a function of small changes in the user's location. 

Lack of measurements poses a challenge to accurate spatially consistent channel modeling, especially for mmWave frequencies. Using field measurements to create and validate the mathematical channel models is one way to ensure accuracy and to gain theoretical insights. The NYUSIM channel model uses realistic large-scale and small-scale parameters for various types of scenarios, environments, and antenna patterns based on massive datasets from measurements at 28, 38, and 73 GHz in urban, rural, and indoor environments \cite{Rap13a,Sun18a,Shafi18a}. Local area measurements were conducted in a street canyon at 73 GHz over a path length of 75 m, where the receiver moved from a non-line-of-sight (NLOS) environment to a line-of-sight (LOS) environment \cite{Rap17b}. The measurements\cite{Rap17b} provide a basis for the proposed model with spatial consistency. 

The NYUSIM channel model simulator operates over a wide range of carrier frequencies from 800 MHz to 100 GHz \cite{Sun17a,Sun18a}, and provides temporal and 3-D spatial parameters for each MPC, and generates accurate CIRs, power delay profiles (PDPs) and 3-D angular power spectrums. The spatial consistency extension proposed here allows the simulator to use additional parameters such as the velocity, location, and moving direction of a user to reproduce realistic CIRs received by the moving user with spatial consistency.  

This paper presents a modified channel coefficient generation procedure for spatial consistency under the framework of the NYUSIM channel model \cite{Ju18a}, and compares the simulation results with 73 GHz measured data from the street canyon measurements \cite{Rap17b}, and is organized as follows. Section \ref{sec:previous} overviews existing models that consider spatial consistency, and provides current approaches for channel tracking. Section \ref{sec:nyusim} describes the impact of spatial consistency on the NYUSIM channel model, and describes the modified generation procedure for spatial consistency. Section \ref{sec:meas} presents the actual channel transitions and resulting CIRs when a user moved in a street canyon based on the measurements. Conclusions are presented in Section \ref{sec:conclusion}. 

\section{Early Research on Spatial Consistency}\label{sec:previous}
Due to the requirements of mmWave communications in mobile and vehicle-to-vehicle (V2V) communications \cite{Perfecto17a}, modern channel models and simulation techniques must adequately characterize changing environments, and generate continuous channel realizations with statistics that are lifelike and usable for accurate simulation for beamforming and other MAC and PHY level design. Channels can be categorized as stationary channels or non-stationary channels based on the rate of the change of the propagation scenarios. Channel modeling and simulations for non-stationary channels where the scattering environment changed significantly, is studied in \cite{Parra18a}. This channel model and simulation method emulated the time-variant nature of a real channel, and realized channel variations in a single channel realization. The channel modeling approach in \cite{Parra18a} can also be extended to stationary channels where the channel parameters are renewed over time while still fulfilling the stationary condition \cite{Parra18a}. Spatial consistency represents the smooth variations for the stationary channels when a user moves, or when multiple users are located closely, in a local area over 5-10 m. 

At microwave frequencies, early statistical CIR models for correlated multipath component amplitudes over one meter local areas due to the small-scale movement were developed from 1.3 GHz measurements, and associated channel simulators, SIRCIM/SMRCIM, were developed based on this model considering spatial and temporal correlation \cite{Rap91b}. Specifically, the simulators considered the motion, the corresponding Doppler spread, and the resulting phase shift on individual multipath components over a local area \cite{Nuckols99,Rap93a}. SIRCIM/SMRCIM simulators were implementations of spatial consistency, before the term was even coined.  

Generally, the small-scale spatial autocorrelation coefficient of the received signal voltage amplitude decreases rapidly over distance, and the correlation distance of individual MPC amplitude is only a few to a few tens of wavelengths. The correlation distance of the received signal voltage amplitude in a wideband (1 GHz) transmission is only 0.67-33.3 wavelengths (0.27-13.6 cm) at 73 GHz, depending on antenna pointing angle with respect to scattering objects \cite{Sun17a,Rap17a}. Furthermore, the amplitudes of individual MPCs in an 800 MHz bandwidth decorrelate over 2 and 5 wavelengths (2.14 and 5.35 cm) at 28 GHz in LOS and NLOS environments, respectively \cite{Samimi16b,Samimi16c}. Spatial consistency, however, is different from small-scale spatial correlation. Spatial consistency refers to the similar and correlated scattering environments that are characterized by large-scale and small-scale parameters in the channel model \cite{Ju18a,Shafi18a}. The large-scale parameters have a much longer correlation distance of 12-15 m \cite{3GPP.38.901} since the scattering environment does not change dramatically in a local area. Small-scale fading measurements \cite{Rap17a} support the hypothesis of spatial consistency extending well beyond one meter since the amplitude of individual MPC, and the total received power varied smoothly and continuously over 0.35 m (the longest distance measured) \cite{Samimi16b}.

Both statistical and deterministic channel models need to be spatially consistent for use in studying adaptive signal processing for mobile scenarios. Statistical channel models rely on large-scale parameters (shadow fading, the number of time clusters, the number of spatial lobes, delay spread, and angular spread) and small-scale parameters (time excess delay, power, AOA and AOD for each MPC) from measurements \cite{Shafi18a}, whereas deterministic channel models rely on geometry and ray-tracing techniques to acquire the channel information \cite{METIS15a,MiWeba14a}.
\begin{figure}[]
    \setlength{\abovecaptionskip}{0cm}
    \setlength{\belowcaptionskip}{-0.5cm}
    \centering
    \includegraphics[width=0.4\textwidth]{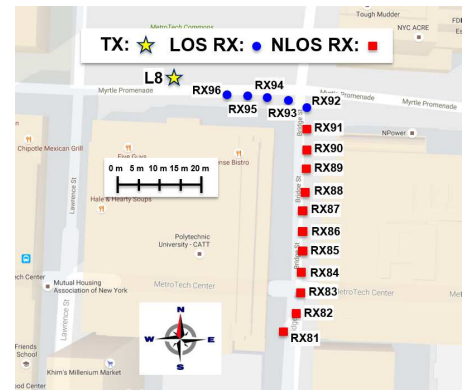}
    \caption{2-D map of TX and RX locations for local area measurements at 73 GHz in a UMi street canyon environment in downtown Brooklyn \cite{Rap17b}. The yellow star is the TX location, blue dots represent LOS RX locations, and red squares indicate NLOS RX locations. North represents 0$^\circ$.}
    \label{fig:route}
\end{figure}
\subsection{Deterministic Channel Models with Spatial Consistency}
Spatial consistency is easier defined and maintained in deterministic channel models, since the locations of the scatterers in the environment are identified in the site-specific channel models \cite{A5GCM15}. The powers, angles, and delays of MPCs can be easily calculated from the relative change of locations of the RX and scatterers based on geometry, generally through the use of ray tracing \cite{Seidel94a}. 

The MiWEBA channel model \cite{MiWeba14a} is quasi-deterministic at 60 GHz, and uses a few strong MPCs obtained from ray-tracing techniques. Several relatively weak statistical MPCs are added to ray-tracing results to maintain some randomness in the channel. The METIS map-based channel model \cite{METIS15a} is also a channel model that uses ray-tracing techniques to acquire large-scale parameters for a specific environment, and uses the map-based large-scale parameters, along with measurement-based statistical small-scale parameters \cite{METIS15a}.
\begin{figure*}[]
    \setlength{\abovecaptionskip}{-1cm}
    \setlength{\belowcaptionskip}{-0.5cm}
    \centering
    \includegraphics[width=0.9\textwidth]{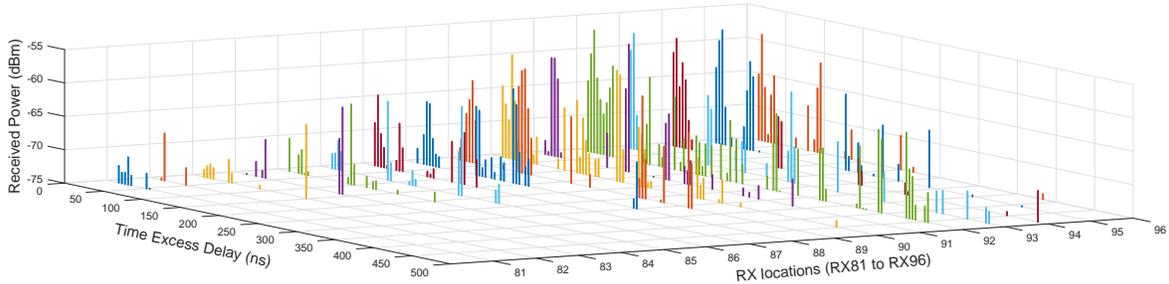}
    \caption{Omnidirectional PDPs at 16 RX locations in a UMi street canyon in downtown Brooklyn at 73 GHz. Referring to Fig. \ref{fig:route}, the receiver moved from RX81(`1' at the `RX locations' axis) to RX96 (`16' at the `RX locations' axis). The distance between two successive RX locations was 5 m. The T-R separation distance varied from 81.5 m to 29.6 m. The visibility condition changed at RX 91 and RX 92 from NLOS to LOS. Absolute time delays are removed to show the difference of time excess delay and delay spread \cite{Rap17b}.}
    \label{fig:omni_16}
\end{figure*}
\subsection{Statistical Channel Models with Spatial Consistency} 
For statistical (e.g. stochastic) channel models, spatial consistency is a challenge since they tend to be drop-based, and cannot generate a time-evolved CIRs in a local area. Thus, geometric information and correlation statistics are necessary for these models to obtain the proper correlated values of large-scale and small-scale parameters for closely spaced locations. 

5GCM \cite{A5GCM15} proposed three approaches for spatial consistency. The first approach uses spatially correlated random variables to generate small-scale parameters such as excess delays, powers, and angles. Users located nearby share correlated values of small-scale parameters. Four complex Gaussian identically and independently distributed (i.i.d) random variables on four vertices of a grid having a side length equal to the correlation distance are generated first. Then, spatially consistent uniform random variables at any location within the grid are formed by interpolating from these four Gaussian random variables \cite{A5GCM15}. The problem with this method of ensuring spatial consistency is that the system needs to store the values of random variables for grids around the user in advance, and requires a large storage space \cite{mmMAGIC17a}. 

The second approach is the geometric stochastic approach \cite{A5GCM15}. In this approach, large-scale parameters are pre-computed for each grid having a side length equal to the correlation distance of the corresponding large-scale parameter. The small-scale parameters are dynamically evolved both in the temporal and spatial domain, based on the time-variant angle of arrivals (AOAs) and angle of departures (AODs), and cluster birth and death \cite{Wang16a}. 

The third approach, the grid-based geometric stochastic channel model (GGSCM) \cite{A5GCM15}, uses the geometric locations of scatterers (i.e., clusters). A cluster is defined as a group of rays coming from the same scatterer, and these rays have similar angles and delays. The angles and delays of the cluster and multipath components in the cluster can be translated into the geometrical positions of the corresponding scatterers. Thus, the time evolution of angles and delays can be straightforwardly computed from the relative changes of the user position, and have very realistic variations. 

MmMAGIC channel models have adopted the three aforementioned spatial consistency approaches, and set the first approach as default since this approach has a more accurate realization of the mmWave channel \cite{mmMAGIC17a}. 

The COST 2100 model is also a geometry-based stochastic channel model \cite{COST2100}, and introduces a critical concept, the \textit{visibility region}. The visibility region refers to a region both in time and space where a group of multipath components is visible to the user. The multipath components in the visibility region constitute the CIRs experienced by the user.

\section{Spatial Consistency Extension for NYUSIM Channel Model} \label{sec:nyusim}

As discussed earlier and in \cite{Ju18a,Sun18a,Shafi18a}, the large-scale and small-scale parameters should vary continuously as a function of the user location in a channel realization over a local area. Under the framework of the NYUSIM channel model \cite{Samimi16a}, a spatial consistency extension is proposed for the NYUSIM channel model and associated simulator \cite{Ju18a}. A spatial exponential filter is applied to make large-scale parameters spatially correlated within the correlation distance of these parameters. The modeling of time-variant small-scale parameters is motivated by the stochastic geometry approach \cite{Wang16a} and the CIR generation procedure in 3GPP Release 14 \cite{3GPP.38.901}. The large-scale path loss is made time-variant, and the shadow fading is made spatially consistent over a local area. Thus, the NYUSIM channel model is extended from a static drop-based to a dynamic time-variant channel model which fits well the natural evolution of NYUSIM and other drop-based statistical models. 

Two distances should be clarified first, the \textit{correlation distance} and \textit{update distance}. The correlation distance determines the size of the grid that maintains spatial consistency of channel conditions. The CIRs of the user moving beyond the correlation distance, or multiple users separated beyond correlation distance, can be regarded as independent. Each large-scale parameter has its own particular correlation distance, and the correlation distance varies according to scenarios and frequencies. For example, the correlation distance of a large-scale parameter in  the UMi scenario is shorter than the one in the RMa scenario because of the higher building density. Thus, extensive propagation measurements for various scenarios and frequencies are necessary to provide the accurate values of the correlation distances of large-scale parameters.

3GPP Release 14 \cite{3GPP.38.901} provides that the correlation distances of large-scale parameters in the LOS and NLOS UMi scenarios were 12 m and 15 m, respectively. Some 73 GHz measurements in a LOS street canyon scenario suggest that the correlation distance of large-scale parameters at 73 GHz is 3-5 m \cite{Wang16a}. From the local area measurements at 73 GHz that will be introduced in the Sec. \ref{sec:meas}, the correlation distance of the number of time clusters is 5-10 m. 
\begin{table}
\centering
\caption{\textsc{Hardware Specifications of Local Area Measurements}}
\label{tab:sys}
\begin{tabular}{|c|c|}
\hline
\textbf{Campaign} & 73 GHz Local Area Measurements  \\
\hline
\textbf{Transmit Signal} & 11$^{\text{th}}$ order PN sequence (length of 2047) \\
\hline
\textbf{TX Antenna Setting} & 27 dBi horn antenna with 7$^\circ$ HPBW \\
\hline
\textbf{RX Antenna Setting} & 20 dBi horn antenna with 15$^\circ$ HPBW \\
\hline
\textbf{TX/RX Chip Rate} & 500 Mcps/499.9375 Mcps \\
\hline
\textbf{RF Null-to-Null Bandwidth} & 1 GHz \\
\hline
\textbf{TX/RX Intermediate Freq.} & 5.625 GHz \\
\hline
\textbf{TX/RX Local Oscillator} & 67.875 GHz (22.635 GHz $\times$ 3) \\
\hline
\textbf{RX ADC Sampling Rate} & 2.5 Msamples/s \\
\hline
\textbf{Carrier Freq.} & 73.5 GHz \\
\hline
\textbf{Max TX Power/EIRP} & 14.3 dBm/41.3 dBm \\
\hline
\textbf{TX-RX Antenna Pol.} & vertical-to-vertical \\
\hline
\textbf{Max Measurable Path Loss} & 180 dB \\
\hline
\end{tabular}
\end{table}

The update distance is the time interval in which the model updates the small-scale parameters for MPCs and renews a CIR. Since the small-scale parameters are time-variant and not grid-based, the update distance should be much shorter than the correlation distance of large-scale parameters to ensure accurate modeling while sampling at an arbitrary time or distance within a local area \cite{Nuckols99}. 3GPP Release 14 suggested that the update distance should be within 1 m. During each update, the channel can be considered as static. That is to say; the update period is 2 s when the user moves at 0.5 m/s; the update period is 0.2 s when the user moves at 5 m/s. 1 m is set as update distance in the NYUSIM channel model for simplicity. 

The details for the generation method of each large-scale and small-scale parameter are described below. 
\begin{itemize}
\begin{figure}[]
    \setlength{\abovecaptionskip}{0cm}
    \setlength{\belowcaptionskip}{-0.5cm}
    \centering
    \includegraphics[width=0.5\textwidth]{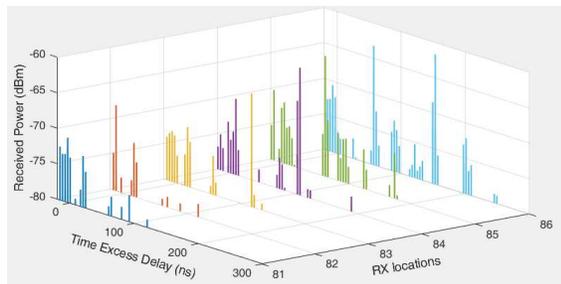}
    \caption{Omnidirectional PDPs at RX81-RX86 to study the correlation distance of large-scale parameters. The distance between two successive RX locations was 5 m. The PDPs at RX81 and RX82 are similar; the PDPs at RX83 and RX84 are similar; the PDPs at RX85 and RX86 are similar. }
    \label{fig:816}
\end{figure}
\item \textit{Time-variant path loss:} The path loss varies smoothly as the user moves in a local area since the shadow fading is spatially consistent. The path loss is obtained from the close-in (CI) path loss model with 1 m free space reference distance \cite{Sun18a}, and calculated in every update period based on the locations of the moving user. The path loss and shadow fading is critical to the evaluation of massive MIMO and multi-user MIMO system performance, and has a large impact on the received power. 
\item \textit{LOS/NLOS transition:} LOS/NLOS condition (LOS probability) determines the value of path loss exponent and shadow fading. Thus, the path loss in LOS and NLOS scenarios are much different. The NYUSIM channel model, as a stochastic model, models the LOS probability as a distance squared model \cite{Rap17a}. The conventional NYUSIM channel model generates the LOS/NLOS condition independently in each simulation. The LOS/NLOS does not change during each simulation. A spatial exponential is applied to make the LOS/NLOS condition spatially correlated based on the correlation distance of LOS probability \cite{Ju18a}. Furthermore, when the LOS/NLOS condition changes, the values of the corresponding parameters will change during the simulation. For Monte Carlo simulations, a statistical spatially correlated LOS/NLOS condition map for a local area is reasonable enough to evaluate the system capacity. However, in the real-world transmission, the information of LOS/NLOS condition in the channel state information (CSI) would be very important for the base station to decide the transmission scheme.  
\item \textit{The number of time clusters, the number of spatial lobes, the number of MPCs in each time cluster:} These three parameters are large-scale parameters that are pre-computed for each grid since the surrounding scatterers do not change rapidly within the correlation distance of large-scale parameters. 
\item \textit{Cluster birth and death:} This concept, first presented in \cite{Wang16a}, demonstrates the time evolution of time clusters. 

When the user moves across grids from location A to location B in the real world, the time clusters appear at the location A may disappear at the location B since the clusters observed at A become very weak. The extension for NYSUIM channel model generates grid-based large-scale parameters including the number of time clusters. Thus, the clusters of A should be discarded, and the clusters of B should be generated gradually during the movement. This procedure can be modeled as a Poisson process with a rate of cluster birth and death. The probability of the occurrence of cluster birth and death is denoted as
\begin{equation}
Pr(t)=1-\exp{(-\lambda_c(t-t_0))}
\end{equation}
where $t_0$ is the most recent update time, and $\lambda_c$ is the mean rate of cluster birth and death per second. This rate varies according to the scenarios, and can only be obtained from field measurements. The birth and death always happen to the weakest cluster at the location. If the numbers of clusters in two grids, A and B, are the same, only the replacement from an old to a new cluster will occur. The weakest cluster of A will be replaced by the weakest clusters of B as one moves from grid A to grid B. Note that when the cluster birth and death occurs, only one cluster of A and one cluster of B will be involved. If the number of clusters in the two grids are not the same, the cluster birth or death will occur alone. This gradual replacement of time clusters ensures spatial consistency in the NYUSIM channel model. 
\end{itemize}

\section{Local Area Measurements for Spatial Consistency} \label{sec:meas}
\subsection{Measurement Environment and Procedure}

Local area measurements were conducted at 73 GHz using a null-to-null RF bandwidth of 1 GHz \cite{Rap17b} to study the spatial consistency and provided the reference values of several parameters such as correlation distance of large-scale parameters. Table \ref{tab:sys} provides the specifications of the measurement system \cite{Rap17b}. The measurements were conducted in a street canyon (18 m wide) between 2 and 3 MetroTech Center in downtown Brooklyn, NY. During the measurements, the TX and RX antennas were set to 4.0 m and 1.5 m, respectively, to emulate the heights of an access point and a user terminal, respectively. TX and RX locations are shown in Fig. \ref{fig:route}, where RX moved from the location RX81 to the location RX96 (NLOS to LOS). The T-R separation distance varied from 81.5 m to 29.6 m. Specifically, the T-R separation distance of NLOS locations (RX81 to RX91) varied from 81.5 m to 50.8 m; the T-R separation distance of LOS locations (RX92 to RX 96) varied from 49.1 m to 29.6 m. The distance between two RX locations was 5 m. Note that the TX antenna pointing angle was the direction that resulted in the strongest received power at the starting location: RX81, and was fixed during the measurements. For each TX-RX combination, the RX swept five times in the azimuth plane. Each sweep was 3 min, and the interval between sweeps was 2 min. The RX antenna swept in half power beamwidth (HPBW) step increments (15$^\circ$). A power delay profile (PDP) was recorded at each RX azimuth pointing angle, and the measurements at each location resulted in at most 120 PDPs (some angles did not have a detectable signal above the noise floor). The best RX pointing angle (the direction where the RX got the maximum received power) in the azimuth plane was selected as the starting direction for the RX azimuth sweeps (elevation remained fixed for all RXs), at each RX location measured \cite{Rap17b}.

\subsection{Measurement Data Processing and Analysis}
\begin{table}
\centering
\caption{\textsc{The Number of Time Clusters in the First 6 RX Locations}}
\label{tab:ntc}
\begin{tabular}{|c|c|}
\hline
\# of time clusters & RX locations \\
\hline
3 & 81,82 \\
\hline
4 & 83,84 \\
\hline
6 & 85,86 \\
\hline
\end{tabular}
\end{table}
24 directional PDPs (HPBW step increments in the azimuth plane, 360/15 = 24) \cite{Rap17b} of one sweep at each location were combined to form one omnidirectional PDP to better illustrate spatial consistency. The denoising was done before this synthesis with a threshold of 20 dB below the peak power of each directional PDP. All 16 omnidirectional PDPs were aligned only for illustration purposes, and the time excess delays of these PDPs are shown in the Fig. \ref{fig:omni_16}. As the RX moved towards TX, the received power increased, and the number of time clusters also increased from 1 up to 6. 

To study the correlation distance of large-scale parameters, the PDPs of the first six NLOS RX locations were studied. The PDPs are shown in Fig. \ref{fig:816}. The number of time clusters is abstracted into the Table. \ref{tab:ntc} based on the time-clustering algorithm described in \cite{Samimi16c}. Thus, the correlation distance of the number of time clusters is about 5-10 m; the correlation distance of delay spread is also about 5-10 m. The same results of correlation distance can be found from the rest PDPs at other LOS and NLOS locations in Fig. \ref{fig:omni_16}.

\begin{figure}[]
    \setlength{\abovecaptionskip}{0cm}
    \setlength{\belowcaptionskip}{-0.5cm}
    \centering
    \includegraphics[width=0.5\textwidth]{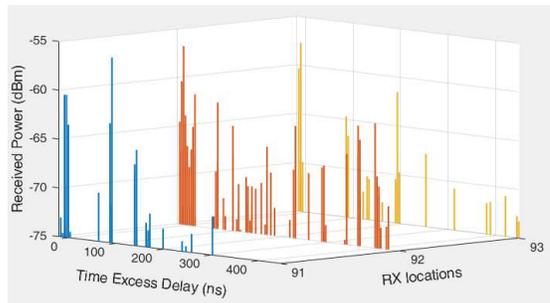}
    \caption{Omnidirectional PDPs at RX91-RX93. The distance between two successive RX locations was 5 m. The receiver at RX 91 was in NLOS condition; The receiver at RX 92 and later locations was in LOS condition.}
    \label{fig:913}
\end{figure}

These local area measurements also showed the impact of LOS/NLOS condition on resulting PDPs. When the RX moved from RX91 to RX92, the visibility condition changed from NLOS to LOS. The PDPs at RX91, RX92, RX93 are shown in Fig. \ref{fig:913}. The received power of RX92 was much stronger than that of RX91, and there were more MPCs at RX92 than at RX91. These results indicate that the LOS/NLOS condition is particularly critical to CIRs and cannot be generated independently for nearby locations as is currently done in conventional statistical channel models. The spatially consistent LOS/NLOS condition would help to predict the CIRs more accurately. 

\section{Conclusion}\label{sec:conclusion}
The spatial consistency extension for the outdoor NYUSIM channel model has been presented in this paper. The generation procedure of both large-scale parameters and small-scale parameters was modified to make these parameters spatially consistent and time-variant. Spatially correlated random variables were applied to characterize the grid-based large-scale parameters; A geometry-based approach was applied to obtain the time-variant small-scale parameters such as time-variant AODs and AOAs, and time cluster birth and death. The static large-scale path loss of drop-based simulations was transformed into a time-variant parameter. The local area measurements in a street canyon were also presented and analyzed in this paper, which indicated that the correlation distance of the number of time clusters and delay spread is about 5-10 m in a UMi street canyon scenario. More field measurements should be conducted to obtain parameters for spatial consistency in various scenarios. The modern channel models with spatial consistency will help in the design of beam tracking and beamforming in system level, and will estimate the channel more accurately for transient simulations.  

\section*{Acknowledgment}
This work is supported in part by the NYU WIRELESS Industrial Affiliates, and in part by the National Science Foundation under Grants: 1702967 and 1731290.

\bibliographystyle{IEEEtran}
\bibliography{gc_copy}

\end{document}